\documentclass[preprint,12pt]{elsarticle}
\usepackage{graphicx}  
\usepackage{dcolumn}   
\usepackage{bm}        
\usepackage{amssymb, nccmath}   
\usepackage{graphicx, epsfig, subfigure}
\graphicspath{{tmpfig/}{figures/}}

\usepackage{amsmath}
\usepackage{slashed}
\usepackage{lipsum}
\usepackage{subfigure}
\usepackage{mathrsfs}
\usepackage{longtable}
\usepackage{rotating}
\usepackage{mathtools}
\usepackage{cuted}
\usepackage{color}
\usepackage{lineno}

\begin{document}

\title{Highly suppressed detection probability of the primordial antimatter in the present-day universe} 


\author[ncku]{Yi Yang} 
\ead{yiyang@ncku.edu.tw}
\cortext[cor1]{Corresponding author.}
\author[as]{Wai Bong Yeung}
\address[ncku]{Department of Physics, National Cheng Kung University, Tainan, 70101, Taiwan, ROC}
\address[as]{Institute of Physics, Academia Sinica, Nankang, Taipei, 11529, Taiwan, ROC}



%


\begin{abstract}
We show that the matter-antimatter asymmetry in the present-day universe is mainly due to the highly suppressed detection probability for the primordial antimatter, which is a direct result of the Dirac-Feynman-Stueckelberg interpretation of antimatter and the extremely time asymmetric expansion of the primordial universe. 
\end{abstract}

  
 

\maketitle


\section{Introduction}
Matter and antimatter are believed to be created in equal amounts in the Big Bang~\cite{big_bang}. 
However, the observed present-day universe is dominated by ordinary matter and this asymmetry between matter and antimatter is a mystery haunting the minds of physicists for many decades. 
In 1967, Andrei Sakharov proposed a framework to fulfill this asymmetry, known as the ``Sakharov conditions''~\cite{Sakharov}, which includes the violation of baryon number, the violation of $C$-symmetry and $CP$-symmetry, and the interactions must be out of thermal equilibrium. 
Hence there are many studies followed along these directions, but unfortunately, until now, none of them can provide conclusive evidence to support it. 
For example, the searches for the $CP$ violating processes in the laboratory end up with effects that are many orders of magnitude too small to explain the asymmetry~\cite{exp_cp_1}\cite{exp_cp_2}\cite{exp_cp_3}. 

Here we propose that the matter-antimatter asymmetry can also be understood as the difference in the detection (scattering) probabilities of primordial matter and antimatter in the present-day universe instead of the different production rates of the primordial matter and antimatter. 
Sakharov proposed his conditions for matter and antimatter asymmetry which were from the point of view of particle physics, and furthermore, the inflationary scenario for the Big Bang was introduced by Alan Guth in 1980~\cite{Guth}, some thirteen years after these conditions were proposed.
So, it is reasonable to assume that the Sakharov conditions may not include enough considerations coming from the expansion or the inflationary expansion of the primordial universe. 
For example, the primordial universe is extremely $T$ asymmetric and there is no guaranteed $CPT$ therom in the primordial universe because the gravitational field was extremely strong at that time which means that non-$T$-invariance may not necessarily follow from $CP$ violation. 
Nevertheless, any attempt to extrapolate the results from the laboratory to the physics in the primordial universe is far-fetched because the laboratory and the primordial universe have different structures of the space-time going with them. 

In this paper, we present an intuitive way which was advocated by Feynman in his space-time approach to quantum mechanics, known as the Feynman propagator approach~\cite{feynman}. 
This is the best tool for investigating the propagation of matter and antimatter and the probability of observing them in a particular specification of the space-time structure.
    We choose to use the Feynman propagator in this paper because it successfully implements the Dirac postulate that the quantum mechanical wave functions of particle propagate forward in time while those of antiparticles propagate backward in time. 
We can then show that the fate of the primordial antimatter is actually determined by the fundamental properties of its wave function and the unique feature of the expanding universe~\cite{Guth}\cite{inflation_2}\cite{inflation_3}, and that the probability of detection of the primordial antimatter in the present-day universe is highly suppressed when compared to that of the primordial matter. 
Most importantly and interestingly, this suppression of the detection of antimatter in the present-day universe is model independent, equally true for primordial positrons, antiprotons, and any other antimatters. 
This suppression stems from the very nature of the space-time behavior of antimatter, the very nature of the extremely rapid expansion of the primordial universe, and the quantum mechanical interpretation of the meaning of detection. 

Our approach of explaining the matter-antimatter asymmetry is organized as following: 
In Section 2, we talk about the hadrons and leptons (antihadron and antileptons) in the universe after its inflationary phase. 
In these post-inflationary epochs, namely the hadron and lepton epochs, the hadrons and leptons were already formed and the universe was still expanding rapidly, but was presumably spatially flat. 
The GUT phase and the Quark-Gluon Plasma phase of the inflationary universe are beyond the scope of this paper. 
A space-time coordinate system in which the Dirac Equation can be implemented will be introduced, and the corresponding propagators for the matter-antimatter will then be formulated. 
In Section 3, these propagators are then calculated by using the Feynman-Stueckelberg prescription, and their explicit forms are derived. 
The first order scattering probability amplitude for the primordial particle and for the antiparticle, respectively, in the present-day universe are derived. 
The scattering probability amplitude for the primordial antiparticle is then shown to be dependent on the scale factor of the universe, and be shown to be highly suppressed when compared with that of the primordial particle. 
It can be concluded, as given in Section 4, that the detection of the primordial antiparticle is highly unlikely, and can be interpreted as its not showing up in the present-day universe. 

\section{Quantum mechanics in the primordial universe}
The interpretation of antimatter in quantum mechanics by Dirac, Feynman, and Stueckelberg can be demonstrated by taking the electron-positron system as an example. 
When Dirac first extended Schr$\ddot{\rm o}$dinger's wave equation into the relativistic domain~\cite{dirac}\cite{bjorken_drell}, he knew that the existence of the negative energy solution can not be ignored and was explained subsequently as a state which carries a positive charge, namely the ``hole'' or the ``positron''. 
Later, the positron state was further interpreted by Feynman and Stueckelberg~\cite{feynman}\cite{stueckelberg} as a state with negative energy which propagates backward in time, while the electron, the positive energy state, propagates forward in time. 
This description of the propagation of electron-positron system is sometime called the Feynman propagator approach~\cite{feynman}. 

\begin{figure}[!htbp]
  \begin{center}
      \includegraphics[width=0.8\textwidth]{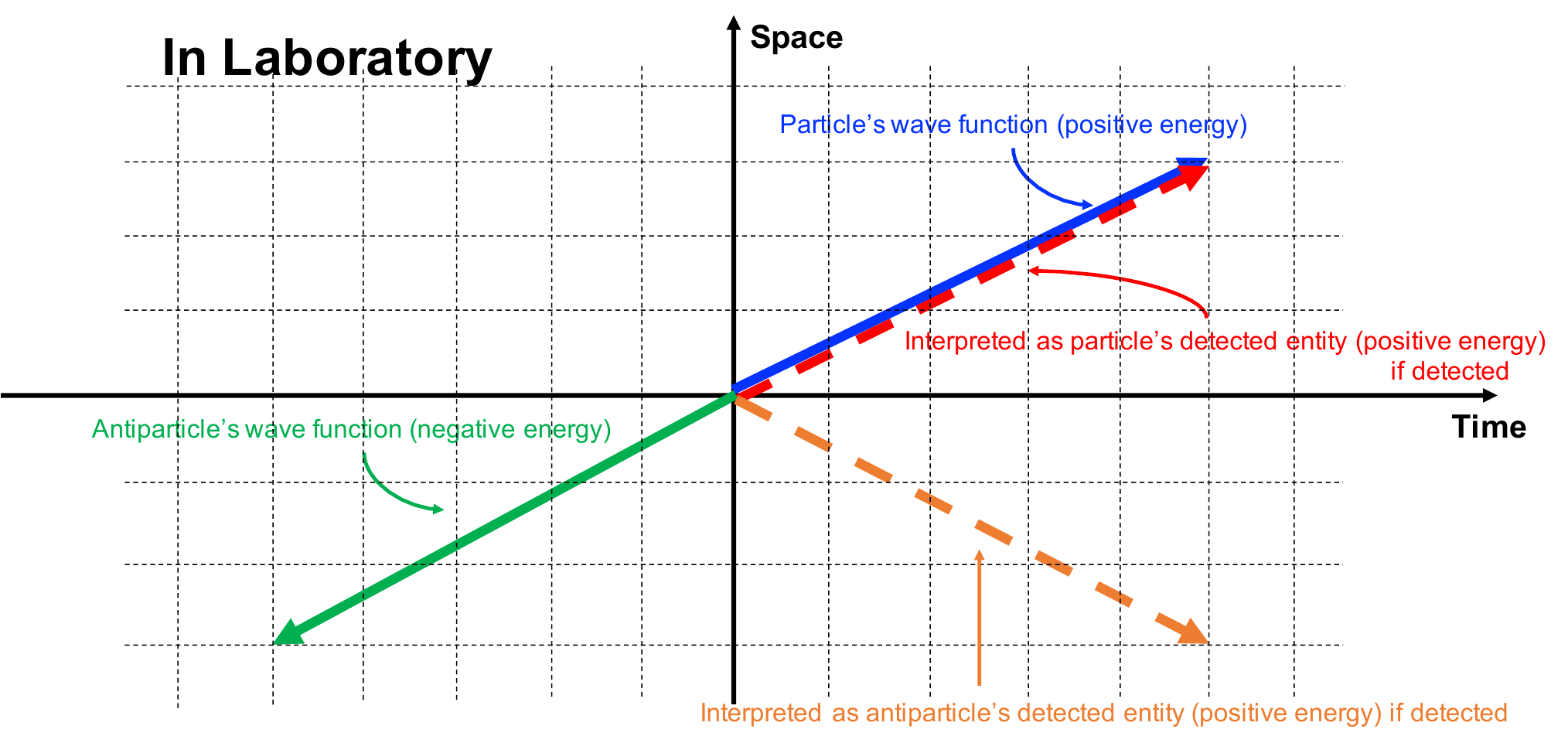}
      \end{center}
    \caption{ The illustration of particle's detected entity (red dashed line) and wave function (blue line) both with interpreted positive energy propagating forward in time, and the antiparticle's detected entity (orange dashed line) with interpreted positive energy propagating forward in time and the wave function (green line) with negative energy propagating backward in time.  
  \label{fig:propagation}}
\end{figure}

While using Feynman's approach, it is important to distinguish carefully between the propagation of the physical entities and the propagation of the corresponding quantum mechanical wave functions. 
The positive energy detected entities (the physical electrons and the physical positrons) are both detected in the forward-time region, while the wave function of the electron with positive energy propagates forward in time and the wave function of the positron with negative energy propagates backward in time, as illustrated in Fig.~\ref{fig:propagation}.  

The statement ``physical positron moves forward in time while its wave function propagates backward in time'' seems self contradictory, however, this is in fact a statement on the quantum properties of the quantum system ``positron''. 
All physical information of the system is contained in its wave function, and the measurements on the wave function tell us the physical behavior of the system.
Here, the detection probability amplitudes of primordial matter and antimatter are determined by the usual quantum mechanical treatment, sandwiching a complex-conjugated wave function and integrating over the entire three-dimensional space, $\int \psi_f^* \psi_i d^3x$ at the time of doing the detection, $i$ here means incoming and $f$ means final.  
If the amplitude is small, then there will be a small chance that the physical primordial electron or positron to be observed in the present-day universe.

In the laboratories, physics looks the same in both time directions.  
However, in the primordial universe, since time is extremely asymmetric, physics will look very different in different time directions. 
It is worthwhile to emphasize that this kind of quantum mechanical wave function propagation with extreme time asymmetry in the primordial universe will hold true for whatever the matter antimatter pairs are. 
In other worlds, the electron and positron pairs will behave the same as the other particle-antiparticle pairs will.  

Next, adding the space-time structure of the primordial universe into our discussions can be started from the Friedmann-Lemaitre-Robertson-Walker (FLRW) metric~\cite{flrw_1}\cite{flrw_2}\cite{flrw_3}\cite{flrw_4} with a cosmic expansion scale $a(T)$ in the cosmic time $T$ in a comoving coordinate system. 
This FLRW metric reads as
\begin{equation}
\label{eq:metric}
    ds^2 = dT^2 - \frac{a(T)^2}{c^2}(dr^2 + r^2d\Omega),  
\end{equation}
where $c$ is the speed of light.
This metric for the cosmos was well documented by the analysis of the CMB temperature fluctuations measured by the Planck Satellite~\cite{planck}. 
These CMB measurements ruled out the hyperbolic or the elliptical spatial geometries for our universe and showed that the observed universe is spatially flat instead. 
As we shall see, this spatial flatness is crucial in our following discussion of the absence of the primordial antimatter in the present-day universe~\cite{pam}.
In other words, the absence of the primordial antimatter can be regarded as a strong support for the spatial flatness of our universe. 
  
\begin{figure}[!htbp]
  \begin{center}
      \includegraphics[width=0.7\textwidth]{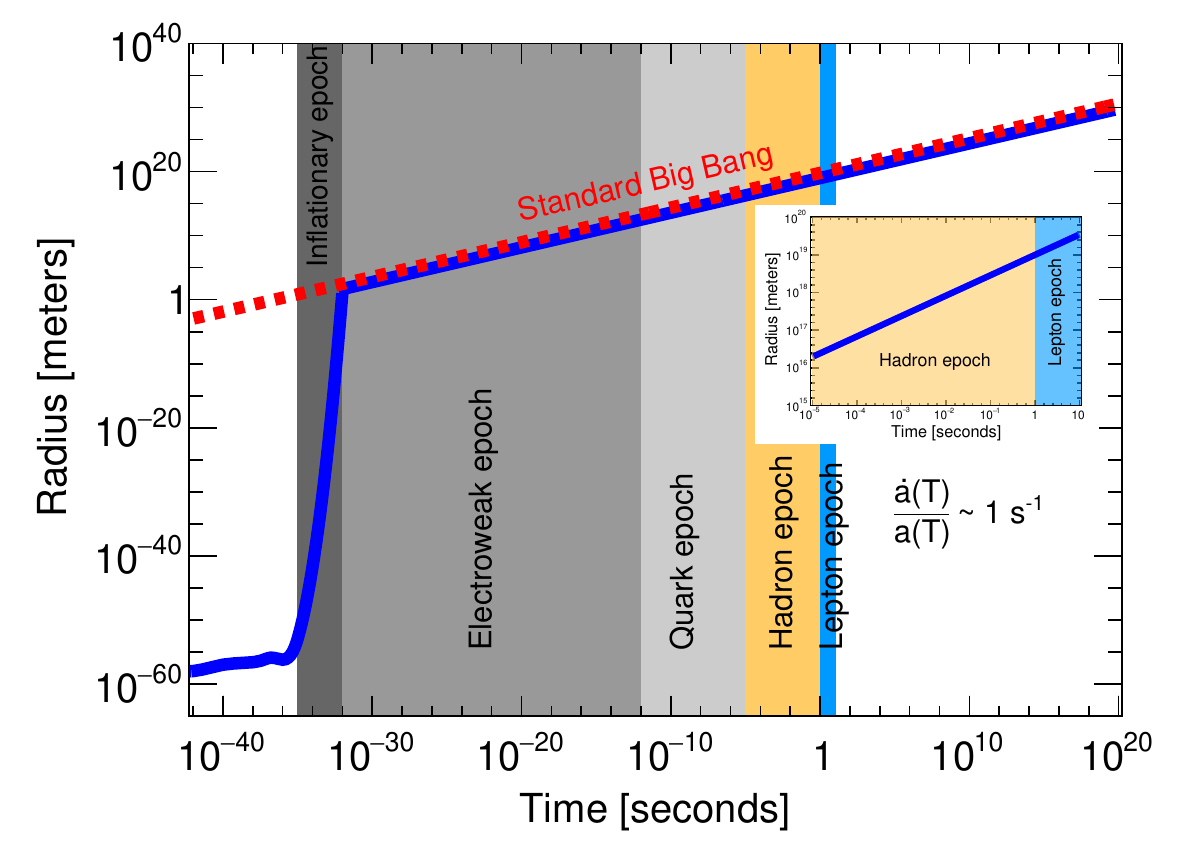}
      \end{center}
    \caption{ The evolution of the radius of the universe. The expansion of the universe with the inflation is shown in the blue line and the standard Big Bang theory is shown in the dashed red line. 
  \label{fig:size}}
\end{figure}

However, the flatness of the FLRW metric is proven only to the CMB freeze out time frame at about 380,000 years after the Big Bang and our knowledge of the metric deeper into the thermal universe is limited. 
Therefore, here we assume that the flat FLRW metric is still valid at the hadron and lepton epochs of the thermal universe.
The hadron and lepton epochs are the time frames when hadrons and leptons dominate the universe. 
Figure~\ref{fig:size} shows the evolution of the radius of the universe.
The physics in these epochs are important for the present-day universe because these hadrons and leptons are precisely what we are now observing. 
The lepton epoch lasted about 10 seconds after the Big Bang~\cite{had_lep_epoch}, and during this period, the radius of the universe changed from about tens of light years to about one hundred light years.
An enlargement of radius by about 10 times during that time frame. 
The expansion rate of the universe at these epochs slowed down a lot when compared with its expansion rate during the inflationary phase, but it was still expanding very fast at this time of the expansion. 
The above facts for the thermal history of the universe tells us that $\frac{\dot{a}(T)}{a(T)}$ = 1 $s^{-1}$ during the lepton epoch. 
Similar argument for the hadron epoch, the radius of the universe was enlarged by 1000 times in $10^5$ seconds, so $\frac{\dot{a}(T)}{a(T)} = 10^{-2}\ s^{-1}$ during the hadron epoch. 
It is important to mention that the expansion rates in the lepton and hadron epochs are taken from the standard thermal history of the cosmos.

    
This expansion mode at this time frame leads us to consider a coordinate system defined as
\begin{eqnarray}
    \label{eq:coord}
    T &=& T, \nonumber \\
    X &=& a(T) r\cos\theta \equiv a(T)x, \nonumber \\
    Y &=& a(T) r\sin\theta\cos\phi \equiv a(T)y,  \\
    Z &=& a(T) r\sin\theta\sin\phi \equiv a(T)z, \nonumber 
\end{eqnarray}
where $r$ is in meters and $T$ is in seconds, while $\theta$ and $\phi$ vary from 0 to $\pi$ and 0 to $2\pi$, respectively. 
The spatial parts of the coordinates will be sometime shortened as $\vec{X} = (X^i) = (X,\ Y,\ Z)$ and $\vec{x} = (x,\ y,\ z)$.
This judicious choice of the coordinate system has the merit of combining the cosmic expansion of the universe with the local motion of the particles (hadrons and leptons) such that $a(T)$ describes the cosmic expansion while $\vec{x}$ describes the local motion. 
Furthermore, $d(X^{i})$ can be approximated as $a(T)dx^{i}$ by using our observation that $\frac{\dot{a}(T)}{a(T)}$ = 1 $s^{-1}$ and $10^{-2}\ s^{-1}$ during the lepton and hadron epoch, respectively, because
\begin{eqnarray}
    \label{eq:dx}
    d(X^{i}) &=& d(a(T))x^{i} + a(T)dx^{i}      \nonumber  \\
             &=& \dot{a}(T) dT x^{i} + a(T)dx^{i},            
\end{eqnarray}
and the relative importance of the two terms on the right-hand side of Eq.~\ref{eq:dx} is given by the ratio of the second term to the first term, which will be denoted by $R$, as
\begin{eqnarray}
    \label{eq:dx_ratio}
    R = \frac{a(T) dx^{i}}{\dot{a}(T)dT x^{i}} = \frac{a(T)}{\dot{a}(T)} \frac{dx^{i}}{dT} \frac{1}{x^{i}}.
\end{eqnarray}
During the hadron and lepton epochs, the universe was presumably still expanding fast but with a much smaller exponential factor of $a(T)_{\rm epochs} = e^{\Lambda_{\rm epochs}t}$ with $\Lambda_{\rm epochs} \approx 1$/sec, and $\frac{dx^{i}}{dT}$ is the local velocity of the particle, which will be very high (close to the speed of light $c$) at such high temperatures. 
Even for the minimum of the above ratio which occurs at $x^{i} = 1$ meter, 
\begin{eqnarray}
    \label{eq:dx_ratio_2}
    R &\approx& \frac{1}{1/{\rm sec}}\ c\ \frac{1}{\rm meter} \approx 3 \times 10^{8}  \ \ {\rm for\ the\ lepton\ epoch} \nonumber \\
      &\approx& \frac{1}{10^{-2}/{\rm sec}}\ c\ \frac{1}{\rm meter} \approx 3 \times 10^{6}  \ \ {\rm for\ the\ hadron\ epoch}.
\end{eqnarray}
So $dX^{i}$ can be well represented by $a(T)dx^{i}$, and 
\begin{eqnarray}
    \label{eq:ds}
    ds^2 &=& dT^2 - \frac{a(T)^2}{c^2}(dr^2 + r^2d\Omega) \\
         &=& dT^2 - \frac{1}{c^2}\left[(dX^1)^2 + (dX^2)^2 + (dX ^3)^2\right]. \nonumber  
\end{eqnarray}
We see that the space-time appears Lorentzian during the hadron and lepton epochs, when using our suggested coordinate system. 
It is important to emphasize that this is just a local effect, valid only in the said epochs.


A crucial benefit of using this coordinate system is that the four momenta for the particles, defined as
\begin{eqnarray}
    P^0 &=& m\frac{dT}{ds} \nonumber \\
    \vec{P} &=& \frac{m}{c} \frac{d\vec{X}}{ds} = m\frac{a(T)}{c}\frac{d\vec{x}}{ds},
\end{eqnarray}
with $m$ as the rest mass of the particle, will satisfy
\begin{equation}
    \label{eq:eng-mom}
    P^{02} - \vec{P}^2 = m^2.
\end{equation}
It is obvious that the redshift of the momenta is already taken care of by the inclusion of the cosmic scale factor $a(T)$ in the definition of $X^{i}$ and the spatial flatness of the universe is crucial in arriving at Eq.~\ref{eq:eng-mom} and the subsequent Feynman propagator which will be described later in Eq.~\ref{eq:prop}.
    
The final assumption is that both quantum mechanics and the Dirac-Feynman-Stueckelberg interpretation of antimatter will work in the hadron and lepton epochs.
The appropriate Dirac Equation can be obtained from linearizing the energy-momentum given in Eq.~\ref{eq:eng-mom} for the hadron-antihadron pairs in the hadron epoch and the lepton-antilepton pairs in the lepton epoch. 
The electron-positron pair will be used as an example in the following discussions.
Therefore, the propagator equation for the electron-positron pair can be written as \cite{bjorken_drell}
\begin{equation}
    \label{eq:prop}
    (i\slashed{\nabla} - m) S_F(X-X') = \delta^4(X-X'),
\end{equation}
where $m$ here is in fact $\frac{mc}{\hbar}$ which is the inverse of the Compton wavelength. 

\section{The detection probabilities for the primordial matter and antimatter in the present-day universe}
\begin{figure}[!htbp]
  \begin{center}
      \includegraphics[width=0.9\textwidth]{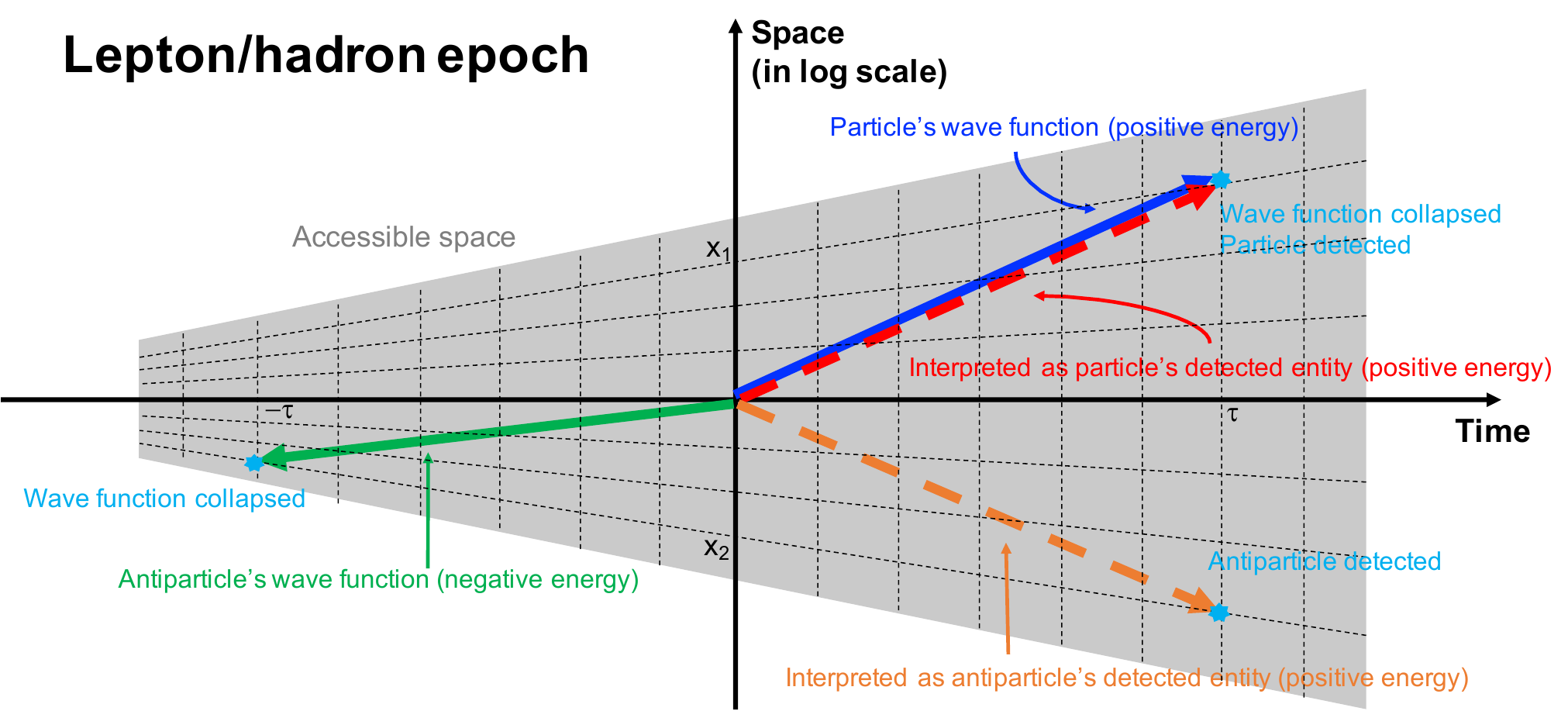}
      \end{center}
    \caption{ The notations are similar as Fig~\ref{fig:propagation}. Gray area indicates the accessible space (in log scale) in the lepton (hadron) epoch. The light blue stars are the places when wave functions for particle and antiparticle collapsed. 
  \label{fig:observation}}
\end{figure}

In the lepton epoch, the temperature is still high enough to produce electron-positron pairs, and once they are produced, as mentioned in the above, the electron wave function will propagate forward in time while the positron wave function will propagate backward in time. 
This statement is also valid for the hadron-antihadron pairs produced in the hadron epoch. 
In the following discussion, the detection of the primordial electron or positron in the present-day universe will be investigated.  
To be precise, the time at $0$ $s$ will be taken as the time when the primordial electron-positron pair was produced and started their propagations in accordance with Eq.~\ref{eq:prop}. 
Figure~\ref{fig:observation} shows an illustration of the electron and positron propagations and detections in the space-time diagram.
The particle's wave function will collapse to a point at $\vec{x}_1$ and $\tau$ when the observation occurred at time $\tau$, while the antiparticle's wave function will collapse at $\vec{x}_2$ and $-\tau$. 
Since the accessible spaces for particle and antiparticle are very different in the lepton (hadron) epoch, the probabilities for observing them will be also very different. 
The detailed calculation will be presented in the following. 

The propagator $S_F(X - X')$ given in Eq.~\ref{eq:prop} represents a wave produced at $X$ by a unit source located at $X'$. 
Here, as mentioned, the source time and the observation time are set to be $0$ $s$ and $\tau$ $s$, respectively, to simplify the notation. 
In other words, 
\begin{eqnarray}
    X' &=& (0,\ \vec{X'}) \equiv (0,\ a(0)\vec{x'}), \nonumber \\
    X  &=& (\tau,\ \vec{X}) \equiv (\tau,\ a(\tau)\vec{x}).
\end{eqnarray}

The solution of Eq.~\ref{eq:prop} for the electron can be obtained by following the Feynman-Stueckelberg approach by adding a small positive imaginary part $+i\epsilon$ to the denominator of the Fourier transform of $S_F$, as following
\begin{equation}
    \label{eq:ele_prop}
    S_F(X-X') = \int \frac{d^4p}{(2\pi)^4} \frac{e^{-ip\cdot(X-X')} }{ p^2 - m^2 + i\epsilon} (\slashed{p} +m),
\end{equation}
where the contour is closed in the lower plane for the $p_0$ integration and the result is
\begin{eqnarray}
    \label{eq:ele_amp_result}
    S_F(X - X')_{\tau>0} &=& -i \int \frac{d^3p}{(2\pi)^3} e^{i\vec{p}\cdot(\vec{X} - \vec{X'})-iE\tau} \left( \frac{E\gamma^0 - \vec{p}\cdot\vec{\gamma} + m}{2E} \right),  \\
     &= & -i \int \frac{d^3p}{(2\pi)^3} e^{i\vec{p}\cdot(a(\tau)\vec{x} - a(0)\vec{x'})-iE\tau} \left(\frac{E\gamma^0 - \vec{p}\cdot\vec{\gamma} + m}{2E} \right), \nonumber
\end{eqnarray}
with $E=+\sqrt{\vec{p}^2 + m^2}$. 

The crucial observation here is that $a(\tau) \gg a(0)$ for a rapidly expanding universe and the $a(0)\vec{x'}$ term can be dropped when compared with the $a(\tau)\vec{x}$ term in the exponential.
Then, this positive energy component becomes 
\begin{eqnarray}
    S_F(X-X')_{\tau>0}= -i \int \frac{d^3p}{(2\pi)^3} e^{i\vec{p}\cdot a(\tau)\vec{x}-iE\tau} \left( \frac{E\gamma^0 - \vec{p}\cdot\vec{\gamma} + m}{2E} \right). 
\end{eqnarray}
After a rescaling of the momentum integration with $a(\tau)\vec{p} = \vec{q}$, the $E$ becomes $\sqrt{\frac{q^2}{a^2(\tau)} + m^2} \approx m$, large values of $q$ can be ignored in the integration, as at such values the exponential oscillates rapidly. 
Then $S_F$ becomes
\begin{eqnarray}
    S_F(X-X')_{\tau > 0} = \frac{-i}{a^3(\tau)} \int \frac{d^3q}{(2\pi)^3} e^{i\vec{q}\cdot\vec{x}-im\tau} \left(\frac{\gamma^0 +1}{2}\right).
\end{eqnarray}

The quantum mechanical detection probability amplitude, ${\mathcal A}^{\rm elec.}$, of the primordial electron in the present-day universe can be calculated by the overlap integral of a incoming plane wave ($e^{ik \cdot X}$) with the wave function of the primordial electron at time $\tau$ which happened after the electron production. 
\begin{eqnarray}
    \label{eq:a_elec}
    {\mathcal A}^{\rm elec.} &=& \int d\vec{X} e^{ik\cdot X} S_F(X-X') \\
       &=& \int a^3(\tau) d\vec{x} e^{+ik_0\tau - i\vec{k}\cdot a(\tau)\vec{x}} \times \left[ \frac{-i}{a^3(\tau)} \int \frac{d^3\vec{q}}{(2\pi)^3} e^{i\vec{q}\cdot\vec{x}-im\tau} \left(\frac{\gamma^0 +1}{2} \right) \right] \nonumber \\
       &=& -i \int d\vec{x} \int \frac{d^3\vec{q}}{(2\pi)^3} e^{+i(k_0\tau - a(\tau)\vec{k}\cdot\vec{x})}  e^{i\vec{q}\cdot\vec{x}-im\tau} \left(\frac{\gamma^0 +1}{2} \right). \nonumber
\end{eqnarray}
Note that some factors in Eq.~\ref{eq:a_elec} are deliberately omitted which are the spinors and they are supposed to be sandwiching the propagator to give the amplitude.
This omission is irrelevant to our result since the spinors are finite even though the momentum is highly shifted because the momentum depending factors which appear in the spinors are $\sqrt{\frac{E+m}{2E}}$ and $\frac{\vec{\sigma}\cdot\vec{p}}{\sqrt{2E(E+m)}}$~\cite{schweber}, which are finite as the incoming momentum $\vec{p}$ goes to infinity,  
It can be seen immediately that the cosmic scale factor cancels out in the amplitude. 
The only modification for the incoming waves is that its momenta is blueshifted by $a(\tau)$ when compared with that observed in the $\vec{x},\ \vec{y},\ \vec{z}$ coordinate system.

For the negative energy component, the story is completely different.
To solve Eq.~\ref{eq:prop} for positrons (or any other antimatter), the contour is closed in the upper plane for the $p_0$ integration, which includes only the negative energy pole.
The source time is now 0 $s$ while the collapse of wave function at $-\tau$ $s$.
Since $a(0) \gg a(-\tau)$, the propagator for the positron can be expressed, this time dropping the $a(-\tau)$ term, as
\begin{eqnarray}
    \label{eq:posi_prop}
     S_F(X-X')_{\tau > 0} = -i \int \frac{d^3\vec{p}}{(2\pi)^3} e^{-i\vec{p}\cdot a(0)\vec{x'}-iE\tau} \times \left(\frac{-E\gamma^0 - \vec{p}\cdot\vec{\gamma} + m}{2E} \right),
\end{eqnarray}
where $X = (-\tau, \vec{X})$ and $X' = (0, \vec{X'})$. 

Again the $a(0)$ factor can be pulled out from the integrand in Eq.~\ref{eq:posi_prop} by rescaling the integral of the momentum, as $a(0)\vec{p} = \vec{q}$.
After convoluting with the incoming wave function $e^{ik\cdot X}$ at time $-\tau$, the detection probability amplitude, ${\mathcal A}^{\rm posi.}$, for the primordial positron becomes
\begin{eqnarray}
    \label{eq:a_posi}
    {\mathcal A}^{\rm posi.} &=& \int d\vec{X} e^{-ik_0\tau - \vec{k}\cdot \vec{X}} S_F(X-X')  \nonumber   \\
    &=&\int a^3(-\tau)d\vec{x} e^{-ik_0\tau - \vec{k}\cdot  a(-\tau) \vec{x}} S_F(X-X')    \\  
    &=& -i \frac{a^3(-\tau)}{a^3(0)} \int d\vec{x} e^{-ik_0\tau - a(-\tau)\vec{k}\cdot\vec{x}}  \nonumber \\
       &&\times \int \frac{d^3q}{(2\pi)^3} e^{-i\vec{q}\cdot \vec{x'}+ \frac{i}{a(0)}\sqrt{\vec{q}^2}\tau}  \left( \frac{-\gamma^0-\vec{q}\cdot\vec{\gamma}/\sqrt{\vec{q}^2}}{2} \right),  \nonumber
\end{eqnarray}
where $X = (-\tau, a(-\tau)\vec{x})$ and $X' = (0, a(0)\vec{x}')$ and the mass term is very small when compared with the $\gamma^0$ term. 
Note that the same omission in Eq.~\ref{eq:a_posi} is made as in Eq.~\ref{eq:a_elec}.
This collapse of quantum mechanical wave function happened at time $-\tau$, a time ``before the appearance'' of the physical positron and the universe was extremely small by that time. 
Interestingly, here, the cosmic factor does appear in the amplitude of the primordial positron for detection in the present-day universe, in the combination of $a^3(-\tau)/a^3(0)$, resulting in a highly suppressed amplitude. 
Note that the incoming wave is now blue-shifted by a factor of $a(-\tau)$, but remains finite when $m \neq 0$~\cite{schweber}.
It is also worthwhile to mention that the integrations in $\vec{x}$ and $\vec{q}$ in Eq.~\ref{eq:a_elec} and Eq.~\ref{eq:a_posi} are in the same order of magnitude which means that the amplitudes for the antiparticle from the primordial universe to be existed in the present-day universe is much smaller than that of for the primordial particle.

The following is a simple example to explain the matter-antimatter asymmetry using our approach. 
For $\tau$ equals to 10 seconds, which is the time span for the lepton epoch, the universe had expanded by 10 times, and $a(-10\ {\rm sec})^3 / a(0\ {\rm sec})^3  = 10^{-3}$. 
Furthermore, the detection probability of the positron at time $\tau$ can be calculated by squaring the ${\mathcal A}^{\rm posi.}$, and the probability is suppressed by a factor of $10^{-6}$. 
This highly suppressed detection probability can be extrapolate to the present-day universe because the scattering amplitude for the electron is independent of the scale factor, and will probably be the major reason of the non-detection of the primordial positron (or any other primordial antimatters) in the present-day universe. 

To summarize, the physical reason for the highly suppressed detection probability for the primordial positron in the present-day universe is due to the fact that its wave function recedes back into the primordial universe which is very small and hence the wave functions of the incident particles will have a very small overlapping chance with the wave function of such primordial positron. 
However the primordial electron will have its wave function fill up the entire present-day universe, and hence has a bigger chance overlapping with the wave functions of the other present-day particles. 
When $\tau$ goes beyond the lepton epoch, the primordial positron will recede further into the hotter universe while the primordial electron will stay in the present-day universe. 
The same statement is true for any hadron-antihadron pairs during the hadron epoch. 
Combining the current $CP$-violation measurements, we may be able to provide the full picture of the non-detection feature of the primordial antimatter, no need for any new mechanisms to provide stronger $CP$-violation. 
Additionally, no effect from quantum gravity is needed and, in fact, we know very little about quantum gravity. 
Finally, the right interpretation of antimatter in accordance with Dirac, Feynman, and Stueckelberg is the key for our success.

We may notice that there are various propagators (Green's functions) appearing in the literature. 
For example, we have the causal propagator, the advanced propagator, and the retarded  propagator, other than the Feynman propagator. 
The causal propagator is defined as the vacuum expectation value of the Poisson bracket of two field operators. 
The advanced propagator is the future part of the causal propagator, while the retarded propagator is the past part of the causal propagator. 
The Feynman propagator is defined as the vacuum expectation value of the time-ordered product of two field operators which, as we know, are related to the scattering amplitudes in quantum field theory. 
And the magnitude of the scattering amplitudes signifies the detection probability of the particles. 
That is why we choose to use the Feynman propagator to calculate the scattering amplitudes to address the detection probabilities of the particles and antiparticles. 
The other propagators find their applications in the discussions of physics of nonequilibrium quantum field theory~\cite{non_eq_qft}.
Of course, one can also use the framework from the quantum field theory in curved space-time or nonequilibrium Quantum Field Theory to study it, but it is, in some senses, way too complicated than using Feynman's propagation approach. 

\section{Conclusions}
Our intuitive argument, by combining the concept of the Dirac-Feynman-Stueckelberg interpretation of antimatter propagating backward in time with the expansion of universe, demonstrates that the detection probability of primordial antimatter in the present-day universe is highly suppressed by a factor of $(a^3(-\tau)/a^3(0))^2$, while the detection probability of the primordial matter is not suppressed. 
Combining with the current $CP$-violation measurements, this may reflect our observation of matter-antimatter asymmetry in the present-day universe. 
Therefore, this paper calls for our attention to the fact that the matter-antimatter asymmetry in the present-day universe maybe determined by the structure of the space-time at the beginning of the universe, and the proper interpretation of antiparticles in the sense of Feynman and Stueckelberg. 

It is important to mention that our proposed approach to look into matter-antimatter asymmetry in the present-day universe can fulfill the three conditions of Sakharov. 
The baryon number violation, as well as the $C$ symmetry and $CP$ symmetry violations, can be understood as the highly suppressed detection probability of the primordial antibaryon in the present-day universe instead of different production rates of baryon and antibaryon. 
In our approach, it is natural that the present-day baryons will outnumber the antibaryons because the wave functions of the antibaryons are receding deep into the even earlier and smaller universe once produced and become hardly detectable by the present-day universe detecting particles. 
In other words, our observable universe has more baryons while the hidden primordial universe has more antibaryons.  
Finally, regarding the condition of the interactions must be out of thermal equilibrium, the wave functions of the particle and antiparticle are going in different time directions and their mutual interactions are getting smaller and smaller once they are produced, and hence there will be a decrease of the occurrence of pair annihilation as the universe expands.
It is also worthwhile to emphasize again that this highly suppressed detection probability of the primordial antimatter in the present-day universe is model independent and is a direct result of the Dirac-Feynman-Stueckelberg interpretation of antimatter and the extremely time asymmetric expansion of the primordial universe.  

\section*{Acknowledgments}
We thank National Cheng Kung University and Academia Sinica for their support. This work was supported in part by the Ministry of Science and Technology of Taiwan and Higher Education Sprout Project from Ministry of Education of Taiwan.

\end{document}